\documentclass[a4paper]{article}
\usepackage{amsmath}
\usepackage{amssymb}
\newcommand{\Mt}{\widetilde{M}} 
\newcommand{\gt}{\tilde{g}}
\newcommand{\scri}{\mathcal{J}}
\newcommand{\HH}{\mathbb{H}}

\newcommand{\TT}{{\mathbb T}}
\newcommand{\VV}{{\mathbb V}}

\renewcommand{\Box}{\square}
\newcommand{\eps}{\varepsilon}

\newtheorem{theorem}{Theorem}
\newtheorem{lem}[theorem]{Lemma}
\newtheorem{prop}[theorem]{Proposition}

\newenvironment{proof}{\begingroup\noindent{\it Proof:\/}}
{\unskip\nobreak\hfil\penalty50
  \hskip2em\hbox{}\nobreak\hfil \vrule depth0pt width6pt height6pt
  \parfillskip=0pt \finalhyphendemerits=0\par\endgroup}

\newcommand{\2}{\tfrac12}

\newcommand{\4}{\tfrac14}

\newcommand{\Psib}{\overline\Psi}
\title{%
Local twistors and the conformal field equations
}
\author{J\"org Frauendiener,\\
Institut f\"ur Theoretische Astrophysik,\\
Universit\"at T\"ubingen,\\
Auf der Morgenstelle 10,\\
D-72076 T\"ubingen,\\
Germany
 \and 
 George A.~J.~Sparling,\\
 Department of Mathematics,\\
 University of Pittsburgh,\\
 Pittsburgh PA, 15260,\\
 USA}

\begin{document}
\maketitle

\begin{abstract}
  This note establishes the connection between Friedrich's conformal
  field equations and the conformally invariant formalism of local
  twistors.
\end{abstract}

The conformal field equations as derived by
Friedrich~\cite{Friedrich-1983} have proved to be a valuable tool in
both analytical and numerical work in general relativity. Not only has
it been possible to derive  global existence theorems for
solutions of Einstein's field equations~\cite{Friedrich-1987,Friedrich-1991},
thereby obtaining rigorous results about the global properties of
asymptotically flat space-times. It was also successfully demonstrated
that the conformal field equations provide a well-defined and very
well-behaved system of equations for numerical purposes
(\cite{Huebner-1996-1,Huebner-1998-2,Huebner-1999-1,jf-1998-1,jf-1997-2,jf-1997-3,jf-1998-2}).

The success of this formulation of the Einstein equations is mainly due
to the fact that the conformal field equations are \emph{conformally
invariant}. This allows for the inclusion of the parts of a space-time
$\Mt$ which are ``at infinity'' with respect to the ``physical
metric'' $\gt$. The extension into the ``unphysical'' space-time
manifold $M$ with metric $g$ is achieved by embedding $\Mt$  into $M$
in such way that on the image in $M$ (which we identify with $\Mt$)
the relation $g = \Omega^2 \gt$ holds, where $\Omega$ is a conformal
factor, i.e., a non-negative scalar function on $M$ with the property
that it is strictly positive on $\Mt$. On the set $\scri := \{ \Omega = 0
\}$ one imposes the condition $d \Omega \ne 0$, by which
$\scri$ becomes a regular three-dimensional hypersurface of $M$. For
more information on this construction we refer to \cite{Penrose-1965}
(see also \cite{jf-1999-2} for a recent review).

The physical metric $\gt$ on $\Mt$ is therefore replaced by a pair
$(g, \Omega)$ on $M$, which gives rise to $\gt=\Omega^{-2} g$ at
points of $M$ with $\Omega \ne 0$, but makes sense also at points
where $\Omega$ vanishes. Obviously, this relationship is not
one-to-one, as there are many pairs $(g, \Omega)$ which give rise to
the same metric $\gt$. In fact, we need to regard $(g, \Omega)$ as
being equivalent to $(\theta^2g, \theta\Omega)$ for any strictly
positive function $\theta$ on $M$. Thus, every equation for $\gt$ has
the property that, when expressed in terms of a pair $(g,\Omega)$, it
is invariant under the rescaling $g \mapsto \theta^2 g$, $\Omega
\mapsto \theta \Omega$. This is the conformal invariance of the
equation in question.

The purpose of this note is to show how the conformal field equations 
of Friedrich can be expressed in the manifestly conformally invariant 
formalism of local twistors~\cite{PenroseRindlerII}.

We start by writing down the conformal field equations. Using the
conventions of~\cite{PenroseRindlerII} throughout, we define $P_{ab} :=
- \2\left( R_{ab} - 4 \Lambda g_{ab}\right)$, where $R_{ab}$ is the
Ricci tensor of the metric $g_{ab}$ with scalar curvature
$R=24\Lambda$. Furthermore, we define $d^a{}_{bcd} := \Omega^{-1}
C^a{}_{bcd}$ and $s := -\4\left(\Box\Omega - 4\Lambda \Omega
\right)$. Note (\cite{Penrose-1965}), that as a consequence of the
smoothness of $\scri$, the Weyl tensor $C^a{}_{bcd}$ vanishes on
$\scri$ so that the ``gravitational field'' $d^a{}_{bcd}$ is regular
there. With these variables the conformal field equations can be
expressed on $M$ as follows:
\begin{gather}
  \nabla_a P_{bc} - \nabla_b P_{ac} = -\nabla_e \Omega d^e{}_{cab},\label{eq:dP}\\
  \nabla_ad^a{}_{bcd} = 0,\label{eq:dd}\\
  \nabla_a \nabla_b \Omega + \Omega P_{ab} + s g_{ab} = 0,\label{eq:dOm}\\
  \nabla_a s - P_{ab} \nabla^a\Omega = 0,\label{eq:ds}\\
  2\Omega s - \nabla_a \Omega \nabla^a \Omega = \lambda/3.\label{eq:OmOm}
\end{gather}
The covariant derivative operator $\nabla$ is the Levi-Civita
connection of the metric $g$.
Several remarks are in order:
\begin{enumerate}
\item These equations are easily checked to be conformally invariant
  in the sense explained above.
\item If one takes $g=\gt$ and $\Omega=1$ then the equations turn out
  to be equivalent on $\Mt$ to the Einstein vacuum equations with
  cosmological constant $\lambda$ together
  with the Bianchi identity for the Weyl tensor.
\item  When these equations are supplemented by the first and second
  Cartan structure equations then one can derive a first order system
  of equations for a tetrad, the connection, the curvature and the
  conformal factor.
\item Upon introduction of suitable ``gauge source functions'' this
  system can be decomposed into a symmetric hyperbolic system 
  of evolution equations and a set of constraint equations. The
  constraints are propagated by the evolution. This is the basis for
  the analytical and numerical applications mentioned above.
\end{enumerate}

We now want to briefly discuss the concept of a local twistor. Let
$(M,g)$ be any (four-dimensional) Lorentzian space-time. Let
$\TT^\alpha(M)$ denote a fibre bundle over $M$ with each fibre
isomorphic to twistor space $\TT^\alpha$. In the usual manner, we may
construct the Grassmann bundle $G_2(\TT^\alpha)(M)$ of two-dimensional
subspaces of $\TT^\alpha$ over $M$. Then the structure of
$\TT^\alpha(M)$ is fixed by the requirement that the fibre of
$G_2(\TT^\alpha)(M)$ over any point be isomorphic to the
(complexified, compactified) Minkowski vector space $T_pM$. This is
nothing but the usual Klein correspondence which allows the
identification of points of Minkowski space with two-dimensional
subspaces of twistor space. Each element of the fibre of
$\TT^\alpha(M)$ over $p$ is called a local twistor at $p$ and a
section $Z^\alpha$ of $\TT^\alpha(M)$ is called a local twistor
(field).

We may identify each tangent vector at $p \in M$ with a
two-dimensional subspace of the fibre of $\TT^\alpha(M)$ over $p$,
which in turn can be identified with a bitwistor $V^{\alpha\beta} = -
V^{\beta\alpha}$ up to scale. Real tangent vectors correspond to
simple bitwistors which satisfy the reality condition
\[
 V_{\alpha\beta} = \overline V_{\alpha\beta}. 
\]
Here, the bar denotes complex conjugation which takes twistors to dual
twistors and $V_{\alpha\beta}=\2\eps_{\alpha\beta\gamma\delta}
V^{\gamma\delta}$ is the dual of $V^{\alpha\beta}$. Also,
$\eps_{\alpha\beta\gamma\delta}$ is the four-dimensional volume on
twistor space.

Each fibre of $\TT^\alpha(M)$ can be considered as a direct sum of two 
two-dimensional spin spaces. However, this decomposition depends on
the conformal scale: if, for a given metric $g$ a local twistor is
represented by $(\omega^A,\pi_{A'})$, then for the conformally related 
metric $\theta^2g$ this same twistor is represented by
$(\omega^A,\pi_{A'} + i\Upsilon_{AA'} \omega^A)$, where 
$\Upsilon_a = \theta^{-1} \nabla_a \theta$. We write 
$Z^\alpha = (\omega^A,\pi_{A'})$, when the metric $g$ is understood. 

There exists a natural connection $D$ on $\TT^\alpha(M)$, the local
twistor transport. It is defined in terms of the representing spinor
fields by
\[
DZ^\alpha = \left( d\omega^A + i\theta^{AA'}\pi_{A'},
  d\pi_{A'} + i P_{ABA'B'}\theta^{BB'} \omega^A \right).
\]
Here, the one-form $\theta^{AA'}$ is the van der Waerden one-form or
soldering form while $P_{ABA'B'}$ is the spinor form of $P_{ab}$
defined above.  It is easily checked that this has the right conformal
transformation properties. The curvature of $D$ can be obtained as
usual from $D^2 Z^\alpha = -iK_\beta{}^\alpha Z^\beta$.

The fact that the tangent spaces of $M$ are not affine spaces but
vector spaces with a preferred origin implies the existence of a
global section $X^{\alpha\beta}$ (unique up to scale), representing
the zero-section of $T(M)$. Equivalently, one can think of
$X^{\alpha\beta}$ as representing the ``current point'' $p$ in the
fibre over $p$.  This ``origin twistor'' is simple, i.e., it satisfies
$X^{\alpha[\beta} X^{\gamma\delta]} = X^{[\alpha\beta}
X^{\gamma\delta]} = 0$.  We have
\[
X^{\alpha\beta} = \left(
\begin{array}{c|c}
0&0\\ \hline 0&\eps_{A'B'}
\end{array}
\right).
\] 
Its covariant derivative is easily computed as
\begin{equation}
  \label{eq:dX}
  DX^{\alpha\beta} = \left(
\begin{array}{c|c}
0&i\theta^A{}_{B'}\\
\hline
-i\theta_{A'}{}^B&0
\end{array}
\right).
\end{equation}
Thus, $DX^{\alpha\beta}$ assumes a r\^ole similar to the soldering
form. As a consequence of~\eqref{eq:dX} we have
\begin{equation}
\label{eq:d2X}
D^2 X^{\alpha\beta} = 2iK_\gamma{}^{[\alpha} X^{\beta]\gamma} = 0.
\end{equation}
which, in view of the previous remark, can be interpreted as stating
that the local twistor connection has no torsion.

From~\eqref{eq:d2X} we conclude that $K_\alpha{}^\beta$ has the form
\[
K_\alpha{}^\beta = \left(
  \begin{array}{c|c}
    i\Psi_A{}^B & \Phi_{AB'} \\
    \hline
    0 & -i \Psib^{A'}{}_{B'}
  \end{array}
\right).
\]
Here $\Psi_{AB} = \Psi_{ABCD} \Sigma^{CD}$ contains only the Weyl
curvature spinor while $\Phi_{AA'} = \nabla_{C(C'} P^C{}_{D')AA'}
\Sigma^{C'D'} + c.c.$ contains only the (derivatives of the) Ricci
curvature. The anti-self-dual two-form $\Sigma^{AB}$ is defined by
$\Sigma^{AB} = \theta^A{}_{A'} \wedge \theta^{BA'}$.

The Einstein equations themselves are not conformally
invariant. Therefore, one needs to introduce an additional structural
element in order to reduce the structure group from the conformal
group to the Poincar\'e group. This is done conventionally by
postulating the existence of an infinity twistor. This is a real
bitwistor (field) $I^{\alpha\beta}$ which is to represent the ``point
at infinity''. Finite points represented by $U^{\alpha\beta}$ have the
property that $U_{\alpha\beta} I^{\alpha\beta} \ne 0$ while for points
at infinity this quantity vanishes. Note, that this interpretation is
valid only if the infinity twistor is simple, which we do not require
here. 

In terms of its representing spinor fields the infinity twistor has
the form
\[
I^{\alpha\beta} = \left(
  \begin{array}{c|c}
    f\eps^{AB} & i\Gamma^A{}_{B'} \\
    \hline
    -i\Gamma_{A'}{}^{B} & g\eps_{A'B'}
  \end{array}
\right)
\]
with real functions $f$, $g$, and a hermitian spinor field
$\Gamma_{AA'}$. Its covariant derivative has the form
\[
DI^{\alpha\beta} = \left(
  \begin{array}{c|c}
    E\eps^{AB} & iE^A{}_{B'} \\
    \hline
    -iE_{A'}{}^{B} & E'\eps_{A'B'}
  \end{array}
\right)
\]
for two scalar one-forms $E$ and $E'$ and a hermitian spinor-valued
one-form $E_{AA'}$. These forms are given by
\begin{align}
  E &= df - \Gamma_{AA'} \theta^{AA'},\label{eq:df}\\
  E_{AA'} &= d\Gamma_{AA'} + g \theta_{AA'} +
  P_{ABA'B'}\theta^{BB'},\label{eq:dGamma}\\
  E' &= dg - P_{ABA'B'}\theta^{BB'}\Gamma^{AA'} \label{eq:dg}.
\end{align}
Since $X^{\alpha\beta}$ represents the current point and since
$X_{\alpha\beta}I^{\alpha\beta}=2f$ we require that $f$ vanishes at
infinity, i.e., at those points where $\Omega=0$. Thus, we make the
ansatz $f=\Omega$. Comparison of equations~(\ref{eq:df}--\ref{eq:dg}) with
the conformal field equations~\eqref{eq:dOm} and \eqref{eq:ds} then shows,
that these equations are exactly equivalent to the equation
\begin{equation}
  \label{eq:dI}
  DI^{\alpha\beta} = 0,
\end{equation}
provided we make the identifications $f\leftarrow \Omega$,
$\Gamma_{AA'} \leftarrow \nabla_{AA'} \Omega$ and $g \leftarrow s$.
As a consequence of equation~\eqref{eq:dI} we have
$
D\left(I^{\alpha\beta}I_{\alpha\beta}\right) = 0
$
and hence the equation 
\[
I^{\alpha\beta}I_{\alpha\beta} = 2\Omega s - \nabla_{AA'}\Omega
\nabla^{AA'}\Omega = \text{const.} = \lambda/3
\]
i.e., equation~\eqref{eq:OmOm}. We note, that the infinity twistor
has the property 
\begin{equation}
  \label{eq:I2}
  I^{\alpha\gamma}I_{\beta\gamma} = \frac{\lambda}{12}
  \delta_\beta^\alpha.  
\end{equation}
Therefore, the infinity twistor is simple if and only if the
cosmological constant vanishes. 

Before we consider the curvature equations~\eqref{eq:dP} and
\eqref{eq:dd} we need to discuss some more properties of the origin
and infinity twistors. We have the
\begin{lem}
\label{lem:proj}
  The two twistors $U_\alpha{}^\beta=X_{\alpha\gamma}I^{\beta\gamma}$
  and $P_\alpha{}^\beta=I_{\alpha\gamma}X^{\beta\gamma}$ possess the
  following properties:
  \begin{description}
  \item[$(i)$] $\displaystyle
    P_\alpha{}^\gamma P_\gamma{}^\beta = \Omega P_\alpha{}^\beta,\qquad
    U_\alpha{}^\gamma U_\gamma{}^\beta = \Omega U_\alpha{}^\beta$,
  \item[$(ii)$] $\displaystyle
    P_\alpha{}^\gamma U_\gamma{}^\beta = U_\alpha{}^\gamma
    P_\gamma{}^\beta = 0,$
  \item[$(iii)$] 
    At points with $\Omega\ne0$ both $\Omega^{-1}P_\alpha{}^\beta$ and
    $\Omega^{-1}U_\alpha{}^\beta$ are projectors onto two-dimensional
    subspaces of $\TT^\alpha$.
  \item[$(iv)$] $\displaystyle
    P_\alpha{}^\beta + U_\alpha{}^\beta = \Omega \delta_\alpha{}^\beta.$
  \end{description}
\end{lem}
\begin{proof}  
The property $(i)$ is immediate when one uses the fact that
$X^{\alpha\beta}$ is simple, which implies
$X_{\alpha[\beta}X_{\gamma]\delta} = -\2 X_{\alpha\delta}
X_{\beta\gamma}$. Property $(ii)$ follows from direct calculation. The
projector property $(iii)$ follows from $(i)$ and the fact that
$P_\alpha{}^\alpha = Q_\alpha{}^\alpha = 2\Omega$. Finally, $(iv)$
follows from $(ii)$ and the four-dimensionality of twistor space.
\end{proof}
These two twistors, $U_\alpha{}^\beta$ and $P_\alpha{}^\beta$, can be
interpreted as projectors onto the unprimed and primed spin
spaces, respectively. 

Next, we consider the curvature equations~\eqref{eq:dP} and
\eqref{eq:dd}. The definition of the gravitational field tensor
$d^a{}_{bcd}$ suggests that we try to divide the curvature
$K_{\alpha}{}^{\beta}$ by the infinity twistor, i.e., to write
$K_{\alpha}{}^{\beta} = D^{\beta}{}_{\alpha\gamma\delta}
I^{\gamma\delta}$ for some unique and regular
$D^{\beta}{}_{\alpha\gamma\delta}$ which has to be determined.  To
begin with we have the following
\begin{lem}
\label{lem:omega}
  Consider the space $\VV_\alpha{}^\beta$ of twistors
  $E_\alpha{}^\beta$, satisfying the equations
  \[ E_\gamma{}^{[\beta} X^{\alpha]\gamma} = 0, \qquad
  E_\gamma{}^{[\beta} I^{\alpha]\gamma} = 0.
  \]
Then the map
\[
\omega: E_\alpha{}^\beta \mapsto I_{\alpha\gamma} E_{\delta}{}^\gamma
X^{\delta\beta} + X_{\alpha\gamma} E_{\delta}{}^\gamma
I^{\delta\beta}
\]
maps $\VV_\alpha{}^\beta$ into itself. Furthermore, it is an
isomorphism at points with $\Omega \ne 0$.
\end{lem}
\begin{proof}
The proof is straightforward once we have made the following
observation. From Lemma~\ref{lem:proj} we obtain for any
$E_\alpha{}^\beta \in \VV_\alpha{}^\beta$
\[
I_{\alpha\gamma} E_{\delta}{}^\gamma X^{\delta\beta} =
I_{\alpha\gamma} E_{\delta}{}^\beta X^{\delta\gamma} =
P_\alpha{}^\delta E_{\delta}{}^\beta
\]
and similarly for the other term, which equals $U_\alpha{}^\delta
E_{\delta}{}^\beta$. Thus, $\omega$ maps $E_\alpha{}^\beta \mapsto \Omega
E_\alpha{}^\beta$ which is obviously an isomorphism if $\Omega\ne0$.  
\end{proof}

Now we find that $\omega$ does exactly we set out to show, namely that 
it allows us to write $K_\alpha{}^\beta$ proportional to the infinity
twistor. For we have the
\begin{prop}
  The curvature $K_\alpha{}^\beta$ of the local twistor connection can 
  be written as 
  \begin{equation}
    \label{eq:Kform}
  K_\alpha{}^\beta = I_{\alpha\gamma} E_{\delta}{}^\gamma
  X^{\delta\beta} + X_{\alpha\gamma} E_{\delta}{}^\gamma
  I^{\delta\beta}
  \end{equation}
  for some uniquely determined hermitian, trace-free twistor
  $E_{\delta}{}^\gamma$. This twistor satisfies the equation
  \begin{equation}
    \label{eq:DE1}
  I_{\alpha\gamma} DE_{\delta}{}^\gamma
  X^{\delta\beta} + X_{\alpha\gamma} DE_{\delta}{}^\gamma
  I^{\delta\beta} + I_{\alpha\gamma} E_{\delta}{}^\gamma
  DX^{\delta\beta} + DX_{\alpha\gamma} E_{\delta}{}^\gamma
  I^{\delta\beta} = 0,
  \end{equation}
  or, equivalently, 
  \begin{equation}
    \label{eq:DE2}
  DE_\alpha{}^\beta + \frac{D\Omega}{\Omega} E_\alpha{}^\beta = 0.
  \end{equation}
\end{prop}
\begin{proof}
  A further consequence of~\eqref{eq:dI} is
\[
  K_{\gamma}{}^{[\alpha} I^{\beta]\gamma} = 0.  
\]
Thus, $K_\alpha{}^\beta$ is a two-form with values in
$\VV_\alpha{}^\beta$ and by lemma~\ref{lem:omega} it is of the
form~\eqref{eq:Kform} at points with $\Omega\ne0$. The fact that
$E_{\delta}{}^\gamma$ is hermitian and trace-free follows from
hermiticity and vanishing trace of $K_\alpha{}^\beta$. We extend this
form for the curvature by continuity to the points with
$\Omega=0$. Then we obtain, that for regular $E_{\delta}{}^\gamma$
some parts of the local twistor curvature must vanish. This is in
complete analogy to the conformal field equations where the regularity
of $d^a{}_{bcd}$ implies that the Weyl curvature vanishes on
$\scri$. The first form of the field equation for $E_\alpha{}^\beta$
follows from the Bianchi identity $DK_\alpha{}^\beta = 0$, while the
second form arises either by simple manipulation of the first, using
the fact that the infinity twistor is covariantly constant, or by
observing that $K_\alpha{}^\beta = \Omega E_\alpha{}^\beta$.
\end{proof}

Now we know that the curvature can be uniquely ``divided'' by the
infinity twistor.  In order to make contact with the conformal field
equations we need to determine the spinor representation of the
equations (\ref{eq:DE1}) or (\ref{eq:DE2}). Denote by
$\HH_\alpha{}^\beta$ the space of hermitian, trace-free twistors
$S_\alpha{}^{\beta}$ which satisfy the equations $S_\gamma{}^{[\beta}
X^{\alpha]\gamma} = 0$ and $S_\gamma{}^{[\beta} I^{\alpha]\gamma} =
0$. It is easy to check that a hermitian $S_\alpha{}^\beta$ with
$S_\gamma{}^{[\beta} X^{\alpha]\gamma} = 0$ is of the form
\[
S_\alpha{}^\beta = \left(
  \begin{array}{c|c}
    S_A{}^B & F_{AB'} \\
    \hline
    0 & \bar{S}^{A'}{}_{B'}
  \end{array}\right)
\]
with a symmetric spinor $S_{AB}$ and a hermitian spinor $F_{AA'}$. It
is obviously trace-free. The further equation $S_\gamma{}^{[\beta}
I^{\alpha]\gamma} = 0$ imposes the additional condition
\[
\Omega F_{AA'} = i \Gamma_A{}^{C'} \bar S_{C'A'} - i \Gamma_{A'}{}^{C}
S_{CA},
\]
which, assuming regularity and defining $S_{AB}=\Omega E_{AB}$, can be
satisfied by writing
\[
S_\alpha{}^\beta = \left(
  \begin{array}{c|c}
    i\Omega E_A{}^B &  \Gamma_{B'}{}^{C} E_{CA} - \Gamma_A{}^{C'} \bar E_{C'B'}\\
    \hline
    0 & -i\Omega \bar{E}^{A'}{}_{B'}
  \end{array}\right).
\]
This is the form that any twistor in $\HH_\alpha{}^\beta$ assumes.  

Since the curvature is a two-form with values in $\HH_\alpha{}^\beta$,
we can write it also in this way, where now $E_{AB}$ and $F_{AA'}$ are
spinor valued two-forms. Stripping off the basis two-forms we obtain
the equations
\begin{gather}
  \Psi_{ABCD} = \Omega E_{ABCD},\\
  \nabla_{C'(C} P^{C'}{}_{D)AA'} = \Gamma_{A'}{}^{B} E_{BACD},
\end{gather}
which are easily verified to be the spinorial equivalents of the
definition of $d^a{}_{bcd}$ and equation~\eqref{eq:dP}.

The remaining equation is obtained from the field equation for
$E_\alpha{}^{\beta}$. After a lengthy calculation one finds that the
only remaining equation which is not identically satisfied due to
earlier equations is
\[
dE_B{}^A - \frac{1}{\Omega} \left( \Gamma_B{}^{B'} \bar E_{B'C'}
\theta^{AC'} - \Gamma_{C'}{}^{C} E_{CB}\theta^{AC'} - \theta^{CC'}
\Gamma_{CC'} E_B{}^A\right) = 0.
\]
Here the first term in the parenthesis vanishes because of the
symmetry of $\bar E_{A'B'C'D'}$, while the remaining two terms are
seen to cancel each other after some manipulation. Thus, the field
equation for $E_\alpha{}^{\beta}$ reduces to the single equation
$dE_B{}^A = 0$, which when written in terms of components is
\[
\nabla_{A'}{}^A E_{ABCD} = 0,
\]
the spinor equivalent of the conformal field equation~\eqref{eq:dd}.

In summary, we have shown the following
\begin{theorem}
  The validity of the conformal field
  equations~(\ref{eq:dP}--\ref{eq:OmOm}) is equivalent to the existence
  of a covariantly constant ``infinity twistor''
  $I^{\alpha\beta}$. The norm of the infinity twistor determines  the
  cosmological constant. In particular, the cosmological constant
  vanishes if and only if the infinity twistor is simple.
\end{theorem}

We should remark that one could have shown the validity of the
equation \eqref{eq:DE2} in an easier way by simply \emph{defining}
$E_\alpha{}^\beta$ by the equation $K_\alpha{}^\beta = \Omega
E_\alpha{}^\beta$ and then using the Bianchi identity for
$K_\alpha{}^\beta$ as before. However, this seems undesirable to us
for two reasons. Firstly, it puts too much emphasis on the conformal
factor $\Omega$, which we want to avoid for reasons discussed
below. And, secondly, it hides the fact, that the division procedure
is an essential consequence of the properties of the two distinguished
bitwistors $X^{\alpha\beta}$ and $I^{\alpha\beta}$.

This concludes our discussion of the conformal field equations in the
version of~\cite{Friedrich-1983}. Our future aim is to also find the
relationship between the newer version of the conformal field
equations and local twistors. In~\cite{Friedrich-1995} it is shown
that one can gain additional freedom by introducing an arbitrary Weyl
connection, which respects the given conformal structure and
expressing the equations in terms of this connection. This removes the
equations \eqref{eq:dOm} and~\eqref{eq:ds} for the conformal factor
$\Omega$, which can therefore be fixed by other methods. The natural
setting for these constructions is the normal conformal Cartan
connection which, as was shown by Friedrich~\cite{Friedrich-1977}
coincides with the local twistor connection. Thus, we expect that
there exists a very natural interpretation in terms of local twistors
also for the more general form of the conformal field equations.

The authors wish to express their gratitude to the Erwin Schr\"odinger
Institute in Vienna and to the Collegium Budapest where part of this
work was made. JF thanks the Deutsche Forschungsgemeinschaft for
financial support.
%\bibliography{gr,jf}
%\bibliographystyle{abbrv}

\end{document}